\documentclass[twocolumn,superscriptaddress]{revtex4}

\usepackage{amsmath}
\usepackage{graphicx}
\usepackage{color}
\usepackage{float}
\usepackage{braket}
\usepackage{amssymb}
\usepackage{bbold}
\usepackage{mathtools}
\allowdisplaybreaks

\newcommand{\idt}{\mathbb{1}}
\DeclareMathOperator{\tord}{\mathcal{T}}
\DeclareMathOperator{\tr}{Tr}

\newcommand{\numb}{\addtocounter{equation}{1}\tag{\theequation}}

\begin{document}

\title{Geometrical Formalism for Dynamically Corrected Gates in Multiqubit Systems}
\author{Donovan Buterakos}
\affiliation{Condensed Matter Theory Center and Joint Quantum Institute, Department of Physics, University of Maryland, College Park, Maryland 20742-4111 USA}
\author{Sankar Das Sarma}
\affiliation{Condensed Matter Theory Center and Joint Quantum Institute, Department of Physics, University of Maryland, College Park, Maryland 20742-4111 USA}
\author{Edwin Barnes}
\affiliation{Department of Physics, Virginia Tech, Blacksburg, Virginia 24061, USA}

\begin{abstract}
The ability to perform gates in multiqubit systems that are robust to noise is of crucial importance for the advancement of quantum information technologies. However, finding control pulses that cancel noise while performing a gate is made difficult by the intractability of the time-dependent Schrodinger equation, especially in multiqubit systems. Here, we show that this issue can be sidestepped by using a formalism in which the cumulative error during a gate is represented geometrically as a curve in a multi-dimensional Euclidean space. Cancellation of noise errors to leading order corresponds to closure of the curve, a condition that can be satisfied without solving the Schrodinger equation. We develop and uncover general properties of this geometric formalism, and derive a recursion relation that maps control fields to curvatures for Hamiltonians of arbitrary dimension. We demonstrate examples by using the geometric method to design dynamically corrected gates for a class of two-qubit Hamiltonians that is relevant for both superconducting transmon qubits and semiconductor spin qubits. We propose this geometric formalism as a general technique for pulse-induced error suppression in quantum computing gate operations.

\end{abstract}

\maketitle

\section{Introduction}

Dynamical gate correction is an important topic in the field of quantum information technology because logical error correction schemes require that the individual gate error be below a given threshold value \cite{FowlerPRA2012}. So, any quantum error correction protocol can only be implemented after individual gate operations are relatively error-free, often necessitating dynamical decoupling of environmental noise by external pulses. Dynamically corrected gates rely on using precise control of the underlying Hamiltonian to perform rotations such that the error introduced by a given noise source at one point in the pulse will exactly cancel the error introduced by the same noise source at other points in the pulse. This idea, adopted from NMR where pulses are used to reduce spin dephasing such as with the Hahn spin echo effect, is used to extend qubit coherence times \cite{HahnPR1950,CarrPR1954,AlexanderRSI1961,ViolaPRA1998,ViolaPRL1999,BrownPRA2004,KhodjastehPRL2005,WitzelPRL2007,WitzelPRB2007,UhrigPRL2007,YangPRL2008,LeePRL2008,UysPRL2009,KhodjastehPRL2010,BiercukJPB2011}. Hahn spin echo and the closely related CPMG type multiple pulsing techniques are essentially the first dynamical decoupling techniques introduced for noise suppression in NMR and ESR experiments. Similarly, pulse sequences have been proposed to perform quantum gate operations robust to various noise sources \cite{KhodjastehPRL2009,vanderSarNAT2012, KhodjastehPRA2012,WangNC2012,WangPRA2014,ThrockmortonPRB2017,ButerakosPRB2018,ButerakosPRB2018_2,EconomouPRB2015,DengPRB2017}. Many of these pulses are derived by finding a series of rotations which combine to produce the desired gate while canceling error to first order or higher. Finding such a sequence often involves numerically solving large systems of nonlinear equations to determine the parameters that define each of the smaller rotations. Solutions to these equations are not unique, and finding one error-correcting pulse sequence only amounts to finding a local maximum in fidelity, and thus many of the resulting pulse sequences are far from optimal. Additionally, many techniques rely on using $\delta$ function or square pulses, neither of which can be precisely implemented in physical systems, as waveform generators have bandwidth and amplitude limitations. Thus, general methods for finding smooth, optimal dynamical-decoupling pulses are highly desirable.

One particularly powerful method for generating single-qubit error-correcting pulses is by representing these pulses as curves parametrized by the cumulative error at any given point in time. Ref. \onlinecite{ZengNJP2018} first showed that for a simple Hamiltonian, pulses that correct against first-order error can be represented as closed curves in a plane, and that the driving field is proportional to the curvature of the curve. Conditions were given for canceling error at higher orders; specifically, for second order, the total signed area enclosed by the curve must equal zero. This formalism was used to derive the fastest possible pulses implementing specific single-qubit gates given constraints on the magnitude of the driving field \cite{ZengPRA2018}. In Ref. \cite{ZengPRA2019}, this formalism was extended to more general single-qubit Hamiltonians with multiple control fields. Here pulses were represented as curves in three dimensions, with the strengths of the driving fields being related to the curvature and torsion of the curve. There have been other proposed techniques for the reverse engineering of the time-dependent Schrodinger equation for dynamical decoupling, but these ideas turned out to be impractical for actual implementation in quantum computing gate operations \cite{FanchiniPRA2007,BarnesPRL2012,BarnesSR2015}.

In this paper, we extend the geometric formalism to systems of multiple qubits subject to quasi-static noise. We show that many of the features of the single-qubit formalism carry over the multiqubit case. Pulses can be represented as curves in a higher-dimensional space, and those which cancel first-order error correspond to closed curves. The length of each pulse is given by the length of the corresponding curve, and the amplitudes of the driving fields are related to the curvature coefficients at each point along the curve. As a demonstration, we derive error-correcting pulses for a class of two-qubit Hamiltonians using curves in six dimensions. We focus on Hamiltonians that describe Ising-type interactions commonly arising in both superconducting transmon qubits and semiconductor spin qubits. We test the performance of these pulses with numerical simulations and confirm that the leading-order infidelity in the presence of noise is canceled.

This paper is organized as follows: in Sec. II, we derive the geometric formalism for a general Hamiltonian. In Sec. III, we give an example of a specific two-qubit system and derive a pulse that implements a single-qubit gate on one qubit while suppressing noise on both qubits. We conclude in Sec. IV.

\section{Multiqubit Geometrical Formalism}

We consider a generic Hamiltonian $H_0(t)$ that contains time-dependent driving fields. Suppose that there is some error term $\delta H$ which is small compared to the scale of $H_0(t)$. In many qubit platforms including superconducting qubits and spin qubits, the time required to perform gates is generally much shorter than the time scale over which $\delta H$ varies \cite{DialPRL2013,OMalleyPRApp2015,MartinsPRL2016,MalinowskiPRL2017,HutchingsPRApp2017}. In this case, even dynamic noise can be well approximated as quasi-static, meaning that the noise fluctuation $\delta H$ is treated as constant during a single pulse, but it can vary from one pulse to the next \cite{WangNC2012}. This is the situation where dynamical decoupling is most effective, and we employ this quasi-static approximation throughout this work. The full Hamiltonian is then $H(t)=H_0(t)+\delta H$. Our goal is to determine driving fields in $H_0(t)$ that perform a desired gate while canceling the effects of noise to leading order in $\delta H$. We consider the evolution operator in the interaction picture. Let $U(t)=\tord e^{-i\int_0^tH(\tau)d\tau}$ be the time-dependent evolution operator of the full noisy Hamiltonian $H$, and let $R(t)=\tord e^{-i\int_0^tH_0(\tau)d\tau}$ be the ideal noiseless evolution operator, where $\tord$ represents the time-ordering operator. This means that the time derivative of $R$ is
\begin{equation}
\frac{dR}{dt}=-iH_0R.
\label{eqn:drdt}
\end{equation}
Then the evolution operator in the interaction picture $U_I(t)$ satisfies $U=RU_I$, and the time-dependent Schr\"odinger equation becomes
\begin{equation}
i\frac{dU_I}{dt}=H_IU_I=(R^\dagger\delta H\;R)U_I.
\end{equation}
The solution to the Schr\"odinger equation to first order in $\delta H$ is given by
\begin{equation}
U_I(t)=\tord e^{-i\int_0^tH_I(\tau)d\tau}=\idt-i\int_0^tH_I(\tau)d\tau+O(\delta H^2).
\label{eqn:ui}
\end{equation}
Note that the integral of $H_I$ is a measure of the total accumulated error at any time $t$, as the gate $U$ actually performed in the presence of noise is equal to the desired gate $R$ only if $U_I=\idt$.

In order to help derive noise-canceling gates, we want to represent this traceless part of $U_I(t)$ as a curve in some $d$-dimensional space. To this end, we define $\mathcal{S}$ to be the vector space on which $H$ acts. For example, if $H$ describes a $n$-qubit system, then $\cal S$ is the Hilbert space of $n$ qubits. The set of all traceless Hermitian matrices which act on $\mathcal{S}$ themselves form another vector space over $\mathbb{R}$, which we will call $\mathcal{V}$ (e.g. $H$ will be an element of $\mathcal{V}$). In the case of $n$ qubits, $\mathcal{V}$ is the space of traceless $n$-qubit Pauli strings, which has dimension $\dim\mathcal{V}=4^n-1$. Define an inner product on this vector space as follows:
\begin{equation}
\vec{V}\cdot\vec{W}=\frac{1}{\dim \mathcal{S}}\tr VW,
\label{eqn:inner}
\end{equation}
where $\dim \mathcal{S}$ is the dimension of the vector space $\mathcal{S}$, and therefore equal to the size of the matrices in $\mathcal{V}$ (e.g. for the case of a system of $n$ qubits, $\dim\mathcal{S}=2^n$). Here, although ${\vec V}= V$, we use the notation $\vec V$, because in what follows, it helps to imagine that we expand $V$ in terms of a basis of Hermitian matrices (e.g., $n$-qubit Pauli strings) with real coefficients. If these coefficients are time-dependent, then they trace out a curve in a higher-dimensional Euclidean space. In what follows, this basis expansion will be kept implicit for the sake of generality and ease of notation. The inner product in Eq. (\ref{eqn:inner}) is invariant under unitary frame transformations, which will become important later.

The possible error terms that can be generated by any specific pulse from a Hamiltonian $H$ will span some subspace of $\mathcal{V}$, which we will call $\mathcal{V}_{\text{err}}$. The subspace $\mathcal{V}_{\text{err}}$ is nonuniversal depending on details, and may or may not be equal to $\mathcal{V}$ depending on the degrees of freedom present in $H_0$ and what rotations are possible. For example, for a single-qubit system, $\mathcal{V}$ is the 3-dimensional space consisting of the $X$, $Y$, and $Z$ Pauli matrices. In Ref. \onlinecite{ZengNJP2018}, it was shown that for a Hamiltonian with a single Pauli matrix, such as $H_0(t)=\Omega(t)X$, the error space $\mathcal{V}_{\text{err}}$ is a 2-dimensional subspace of $\mathcal{V}$. In comparison, Ref. \onlinecite{ZengPRA2019} examined a Hamiltonian with two different driving terms such as $H_0(t)=\Omega_X(t)X+\Omega_Y(t)Y$, and found that $\mathcal{V}_{\text{err}}$ in fact spans the entire space of $\mathcal{V}$ in this case. In general, $\mathcal{V}_{\text{err}}$ is the space spanned by $\delta H$ and all vectors to which $\delta H$ can be rotated by the evolution operator $R$ corresponding to a particular choice of $H_0(t)$. For example, consider a Hamiltonian of the form
\begin{equation}
H_0(t)=\Omega_1(t)V_1+\Omega_2(t)V_2+...+\Omega_k(t)V_k,
\end{equation}
where $V_i$ are products of Pauli matrices, and $k$ is some arbitrary integer. Then $\mathcal{V}_{\text{err}}$ will include linear combinations of $\delta H$ and all vectors of the form $i[V_i,\delta H]$. Additionally, if any terms in $H_0$ do not commute with each other, then the time ordering of $R$ will allow sequences of rotations, and thus nested commutators such as $-[V_j,[V_i,\delta H]]$ will also be included in $\mathcal{V}_{\text{err}}$.

Let $d$ be the dimension of $\mathcal{V}_{\text{err}}$, and define a $d$-dimensional vector $\vec{G}(t)$ as follows:
\begin{align}
\vec{G}(t)=\frac{1}{|\delta H|}\int_0^tH_I(\tau)d\tau.
\end{align}
This gives the accumulated error present in Eq. (\ref{eqn:ui}) as a function of time, divided by $|\delta H|$ so that $\vec{G}(t)$ is independent of the actual strength of the noise. Here, the magnitude is given by $|\vec{A}|=\sqrt{\vec{A}\cdot\vec{A}}$ for any matrix $\vec{A}$. We can picture $\vec{G}(t)$ as tracing out a curve in $\mathcal{V}_{\text{err}}$, with $\vec{G}(0)=0$. If $\vec{G}(t)$ returns to zero at some later time $t_{\text{pulse}}$, then the rotation performed will be unaffected by error to first order in $|\delta H|$, and thus dynamically corrected gates correspond to closed $d$-dimensional curves by construction. Now consider the time derivative of $\vec{G}(t)$:
\begin{equation}
\bigg|\frac{d\vec{G}}{dt}\bigg|=\frac{|H_I|}{|\delta H|}=1.
\end{equation}
Since $d\vec{G}/dt$ always has unit length, the distance along the $d$-dimensional curve parametrized by $\vec{G}(t)$ corresponds to the time $t$ at that point. Thus, the tangent vector to the curve at any point is given by $d\vec{G}/dt$, and the normal vector is proportional to the second time derivative of $\vec{G}$. Using Eq. (\ref{eqn:drdt}), we can express this in terms of $R$, $H_0$, and $\delta H$ as follows:
\begin{align}
\frac{d^2\vec{G}}{dt^2}&=\frac{1}{|\delta H|}\frac{dH_I}{dt}=\frac{1}{|\delta H|}\frac{d}{dt}\big(R^\dagger\;\delta H\; R\big)\nonumber\\
&=\frac{i}{|\delta H|}R^\dagger\big[H_0,\delta H\big]R.
\label{eqn:d2g}
\end{align}
The curvature $\kappa_1$ at any point along the curve can be calculated from this, and is given by
\begin{align}
\kappa_1=\bigg|\frac{d^2\vec{G}}{dt^2}\bigg|=\bigg|\frac{i}{|\delta H|}R^\dagger\big[H_0,\delta H\big]R\bigg|=\bigg|\frac{i}{|\delta H|}\big[H_0,\delta H\big]\bigg|.
\label{eqn:kappa1}
\end{align}

In this representation of the cumulative error $G(t)$ as a $d$-dimensional curve, the ideal evolution operator $R(t)$ is a rotation which takes the initial tangent vector $\delta H$ to the tangent vector $d\vec{G}/dt$ at any time $t$ along the curve. Because $R$ is a time-ordered exponential, it can be difficult to work with, since it cannot be analytically calculated for most choices of $H_0(t)$. However, as Eq. (\ref{eqn:kappa1}) shows, this does not prevent us from calculating the curvature, since the curvature at any point on a curve does not depend on the absolute orientation of the curve, but only on points in its own local neighborhood. This is reflected in Eq. (\ref{eqn:kappa1}) by the fact that $\kappa_1$ does not depend directly on $R$. Thus, for a given $\delta H$, the curvature provides a direct relationship between $G(t)$ and the Hamiltonian $H_0$, without the need to calculate $R$. If $d>2$, the same reasoning demonstrates that the higher-order curvature coefficients are independent of $R$ as well. These can be calculated using the Frenet-Serret equations:
\begin{equation}
\vec{e}_n=\frac{d^n\vec{G}}{dt^n}-\sum_{j=1}^{n-1}\bigg(\frac{d^n\vec{G}}{dt^n}\cdot\hat{e}_j\bigg)\hat{e}_j.
\label{eqn:fsvec}
\end{equation}
Here, the vectors $\vec{e}_n$ (more precisely their normalized counterparts $\hat{e}_n$) define the Frenet-Serret frame at each point along the curve defined by $\vec{G}$. The number of vectors is therefore equal to $d$, the dimension of $\mathcal{V}_{\text{err}}$. In the single-qubit case for example where $d=3$, $\vec{e}_1$, $\vec{e}_2$ and $\vec{e}_3$ are the tangent, normal and binormal vectors, respectively. Knowing how these vectors evolve in time is equivalent to knowing how the curve moves through space. These vectors evolve in an interdependent way that is governed by the generalized curvatures:
\begin{equation}
\kappa_n=\frac{d\hat{e}_n}{dt}\cdot\hat{e}_{n+1}.
\label{eqn:fscurv}
\end{equation}
In the single-qubit case, $\kappa_1$ is the curvature of the three-dimensional space curve, while $\kappa_2$ is its torsion. More generally, $d-1$ curvatures are needed to characterize a curve in $d$ dimensions. The higher-order derivatives of $\vec{G}(t)$ can be calculated in a similar fashion to Eq. (\ref{eqn:d2g}) above, which will generate nested commutators. There will, however, be additional terms due to the time dependence of $H_0$. Define the operator $C$ which acts on a matrix $V$ as follows:
\begin{equation}
C\,V=i[H_0,V].
\end{equation}
Then the derivatives of $\vec{G}$ are given by
\begin{equation}
\frac{d^n\vec{G}}{dt^n}=\frac{1}{|\delta H|}R^\dagger\bigg(\Big(C+\frac{\partial}{\partial t}\Big)^{n-1}\delta H\bigg)R.
\label{eqn:dng}
\end{equation}
It is important to note that $C$ does not commute with $\partial/\partial t$, and thus there will be increasingly more terms in this expression for higher values of $n$. Eqs. (\ref{eqn:fsvec})-(\ref{eqn:dng}) relate the generalized curvatures to the driving fields.

While Eq. (\ref{eqn:dng}) can be difficult to work with in general, there are specific conditions under which it can be simplified. Specifically, suppose that the following anticommutation relation holds: $\{H_0,\delta H\}=0$. Note that if $\delta H$ is proportional to a Pauli string, then one can always transform to a basis in which this relation holds \cite{ZengPRA2019}. For convenience, define $Q=\delta H/|\delta H|$, and assume that $Q^2=1$. Define the operator $A_n$ such that the Frenet-Serret basis vectors can be written as:
\begin{equation}
\hat{e}_n=R^\dagger A_nQR,
\label{eqn:andef}
\end{equation}
Differentiating Eq. (\ref{eqn:andef}) yields the following:
\begin{equation}
\dot{\hat{e}}_n=iR^\dagger\{H_0,A_n\}QR+R^\dagger\dot{A}_nQR.
\label{eqn:de}
\end{equation}
On the other hand, the Frenet-Serret equations read:
\begin{equation}
\dot{\hat{e}}_n=\kappa_n\hat{e}_{n+1}-\kappa_{n-1}\hat{e}_{n-1}.
\label{eqn:fs}
\end{equation}
We now show that the two terms in Eq. (\ref{eqn:de}) directly correspond to the two terms in the Frenet-Serret equations. To do this we will show that $[A_n,Q]=0$ for $n\!\!\mod 4=0$ or $1$, and $\{A_n,Q\}=0$ for $n\!\!\mod 4=2$ or $3$. We will make use of the property that for any operators $O_1$ and $O_2$ such that $O_1$ commutes with $Q$ and $O_2$ anticommutes with $Q$, the following holds:
\begin{equation}
\tr O_1QO_2Q=0.
\label{eqn:oqdot}
\end{equation}
Consider the first Frenet-Serret vector (the tangent vector to the curve):
\begin{equation}
\hat{e}_1=R^\dagger QR.
\end{equation}
Here $A_1=\idt$, and thus $[A_1,Q]=0$ holds trivially. Differentiating $\hat{e}_1$, we obtain the second vector:
\begin{equation}
\dot{\hat{e}}_1=\kappa_1\hat{e}_2=iR^\dagger\{H_0,A_1\}QR,
\end{equation}
from which we find that $\kappa_1A_2=i\{H_0,A_1\}$, and the relationship $\{A_2,Q\}=0$ follows. Similarly, we can consider the derivative of the $n$th vector given by Eq. (\ref{eqn:de}). If $A_n$ commutes with $Q$, then $\dot{A}_n$ will also commute with $Q$, and $\{H_0,A_n\}$ will anticommute with $Q$. Similarly, if $A_n$ anticommutes with $Q$, then $\dot{A}_n$ will anticommute with $Q$, and $\{H_0,A_n\}$ will commute with $Q$. In either case, Eq. (\ref{eqn:oqdot}) implies that the two terms in Eq. (\ref{eqn:de}) will be orthogonal to one another. Additionally, one of these two terms will be orthogonal to $\hat{e}_{n-1}$. This means that terms in Eq. (\ref{eqn:de}) can be matched to the terms in Eq. (\ref{eqn:fs}), and the curvatures and operators $A_n$ can be obtained. Specifically, the following recursion relation holds:
\begin{equation}
\kappa_nA_{n+1}=\begin{cases}
i\{H_0,A_n\}&\text{if $n$ is odd}\\
\dot{A}_n&\text{if $n$ is even}
\end{cases}
\label{eqn:recurs}
\end{equation}
The curvatures are obtained by taking the magnitude of this expression, since $A_n$ has unit magnitude. Eq. (\ref{eqn:recurs}) thus provides a general mapping between control fields in the Hamiltonian and curvature coefficients of the curve.

\section{Deriving Example Pulses}

In this section, we apply the general geometrical formalism derived in Sec. II to a specific two-qubit Hamiltonian, and use it to derive pulses that implement dynamically corrected gates in the presence of quasi-static noise.

Consider the case in which the qubits are coupled by an Ising-type interaction, which creates an energy splitting between the $\ket{00}\leftrightarrow\ket{01}$ and $\ket{10}\leftrightarrow\ket{11}$ transitions. This type of interaction is common in solid-state qubit systems. For example, it arises in the context of superconducting transmon qubits from both direct capacitive coupling and also from resonator-mediated interactions \cite{KochPRA2007,MagesanPRA2020}. It also applies to semiconductor spin qubits, both for exchange-based \cite{WardropPRB2014,QiaoARXIV2020} and capacitive \cite{ShulmanSci2012,WangNPJQI2015,NicholNPJQI2017} inter-qubit couplings. Single-qubit or two-qubit gates can then be performed by driving only one qubit (see e.g., Refs. \onlinecite{EconomouPRB2015,MagesanPRA2020,CalderonVargasPRB2019}). The noiseless Hamiltonian in this case is
\begin{align*}
H_0&=\Omega(t)X_2+\frac{E_1+E_2}{2}Z_2+\frac{E_1-E_2}{2}Z_1Z_2\\
&=\begin{pmatrix}
E_1&\Omega(t)&0&0\\
\Omega(t)&-E_1&0&0\\
0&0&E_2&\Omega(t)\\
0&0&\Omega(t)&-E_2
\end{pmatrix}
\numb
\label{eqn:h0}
\end{align*}
where $X_i$ and $Z_i$ are the Pauli matrices on qubit $i$. Suppose that the leading source of noise $\delta H$ is proportional to $Z_2$, corresponding to slow fluctuations in the energy splittings of both qubits, as may arise, for example, from charge and field noise. Then using our geometrical formalism, we can represent the cumulative error as a curve in the 6-dimensional space $\mathcal{V}_{\text{err}}$ spanned by $\{X_2,Y_2,Z_2,Z_1X_2,Z_1Y_2,Z_1Z_2\}$. Using Eqs. (\ref{eqn:fscurv}) \& (\ref{eqn:dng}), we calculate the five curvature coefficients as follows:
\begin{align*}
\kappa_1&=2|\Omega|,\\
\kappa_2&=\sqrt{2(E_1^2+E_2^2)},\\
\kappa_3&=\frac{\sqrt{2}|E_1^2-E_2^2|}{\sqrt{E_1^2+E_2^2}},\\
\kappa_4&=2\sqrt{\Omega^2+\frac{2E_1^2E_2^2}{E_1^2+E_2^2}},\\
\kappa_5&=\frac{E_1E_2\sqrt{2(E_1^2+E_2^2)}}{\Omega^2(E_1^2+E_2^2)+2E_1^2E_2^2}\;\frac{d\Omega}{dt}.
\numb
\label{eqn:curvature}
\end{align*}

Every pulse $\Omega(t)$ corresponds to some 6-dimensional curve given by these curvature coefficients. It is important to realize that the converse is not true: every 6-dimensional curve does not necessarily correspond to a pulse that can be created by the Hamiltonian given by Eq. (\ref{eqn:h0}). This is because there are in general five degrees of freedom corresponding to a 6-dimensional curve (corresponding to the five curvature coefficients), but in Eq. (\ref{eqn:h0}) we have constrained the form of the Hamiltonian such that it has only one independent driving field. Because the Hamiltonian given by Eq. (\ref{eqn:h0}) is block diagonal, the system can alternatively be treated as two independent $2\times 2$ subsystems corresponding to the two blocks, with the constraint that the driving field $\Omega$ must be the same in each block. These subsystems can each be treated using the single-qubit formalism given in Ref. \onlinecite{ZengPRA2019}. This means that a closed 6-dimensional curve which satisfies Eq. (\ref{eqn:curvature}) can be mapped to two separate 3-dimensional closed curves with equal lengths and curvatures to each other, but different torsions given by $E_1$ and $E_2$.

\begin{figure}[!tb]
	\includegraphics[width=.6\columnwidth]{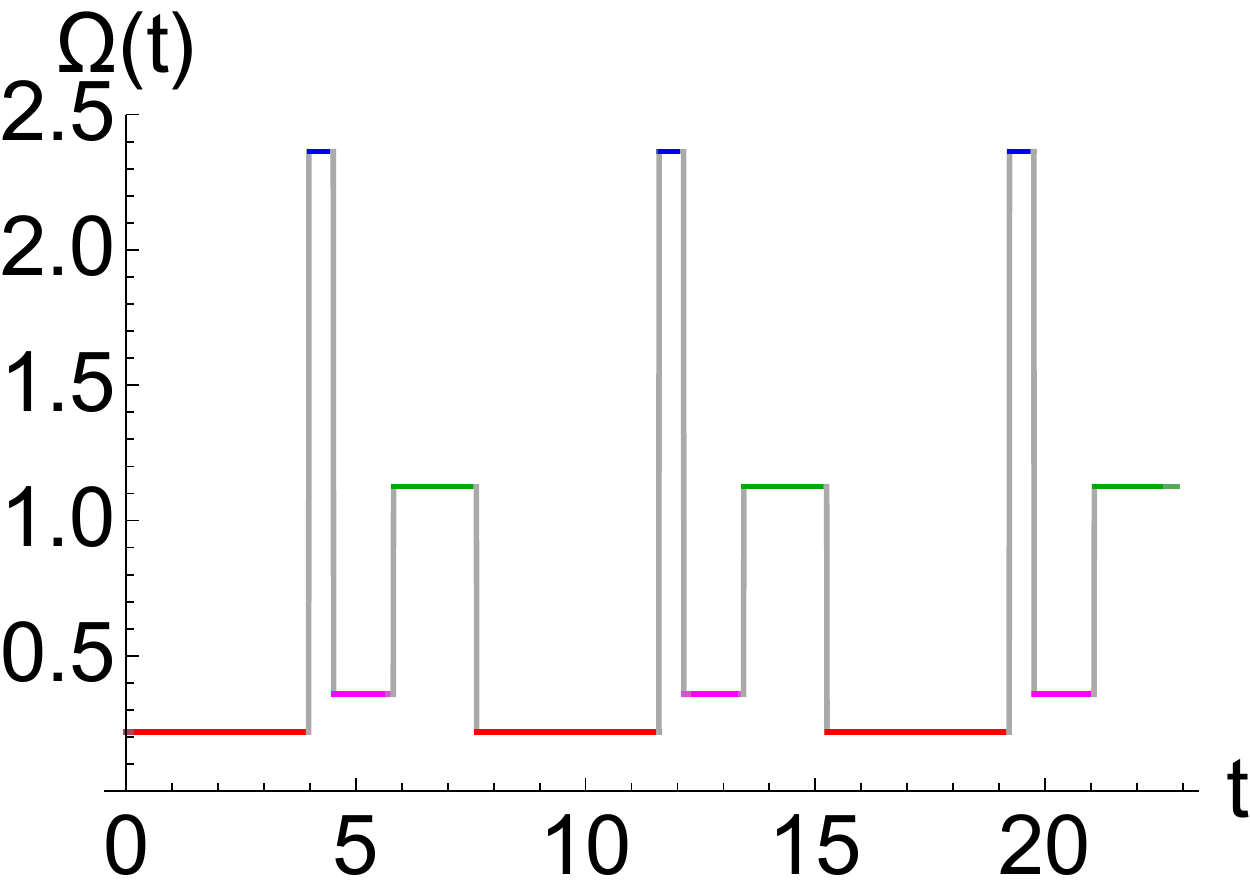}
	\includegraphics[width=.49\columnwidth]{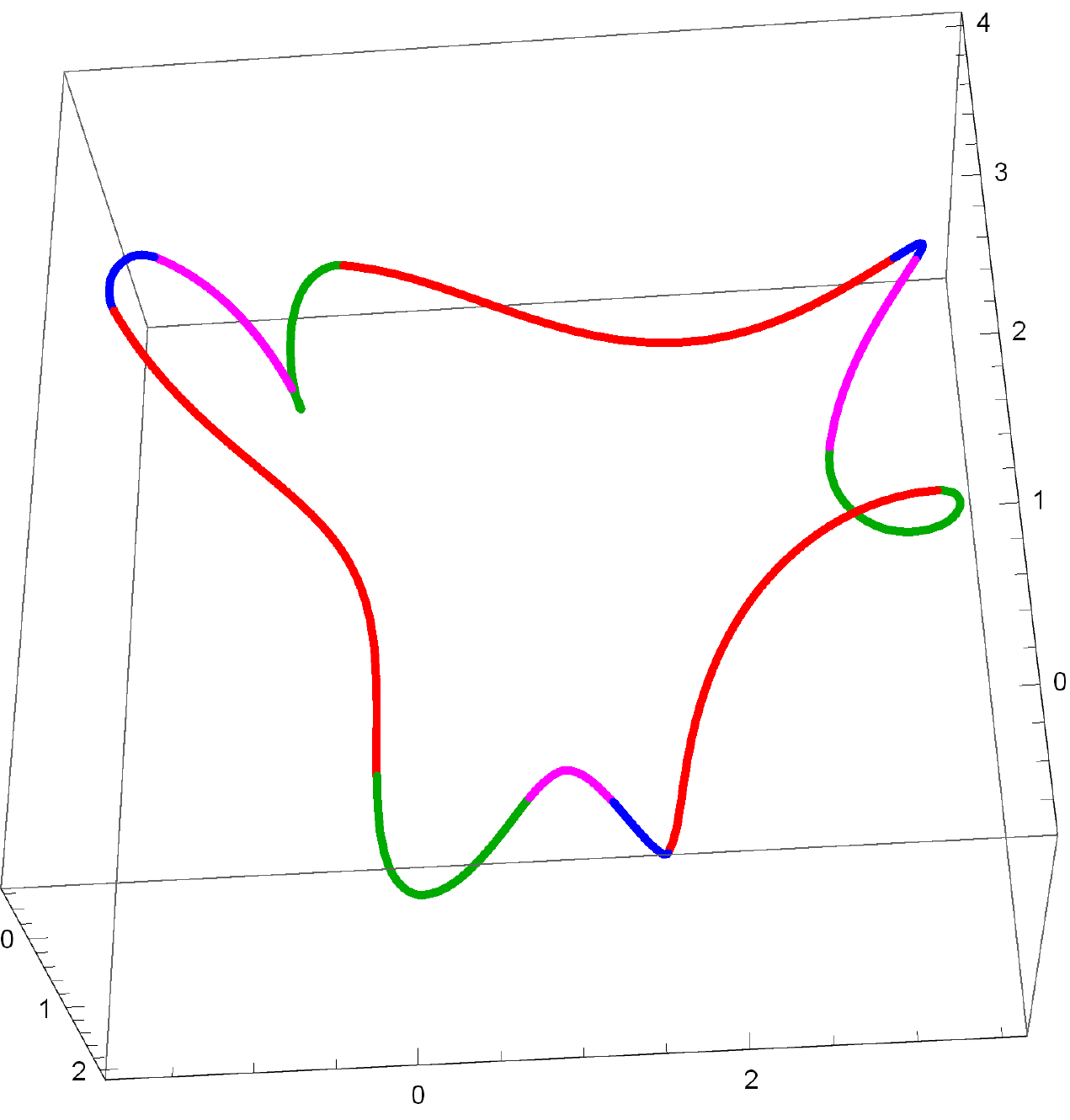}
	\includegraphics[width=.49\columnwidth]{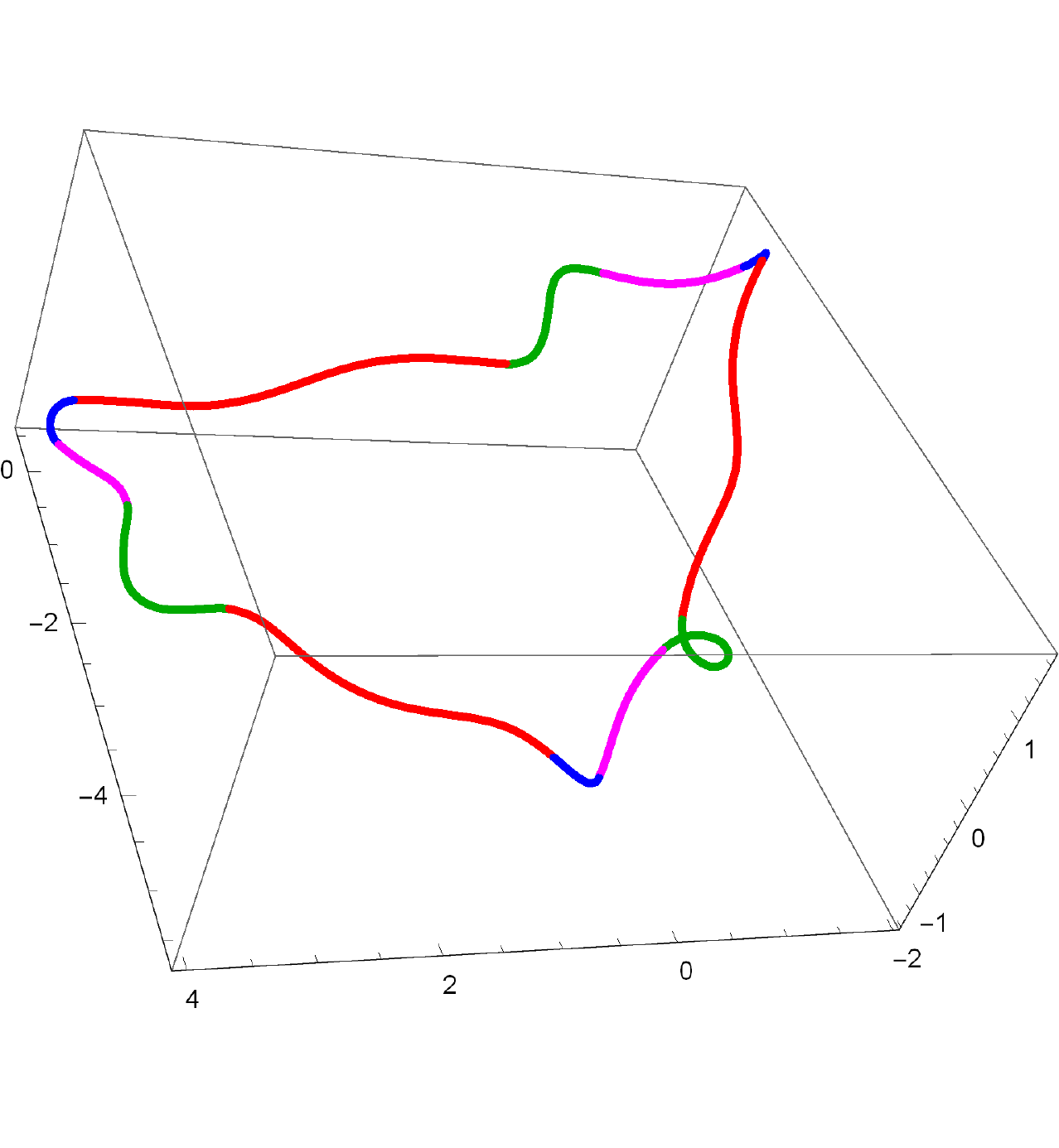}
	\caption{{\bf Top:} A square pulse sequence implementing a $Z_1$ gate which dynamically cancels error to first order in $\delta H$ for the Hamiltonian given by Eq. (\ref{eqn:h0}). {\bf Bottom:} Error curves for the two $2\times 2$ blocks of the Hamiltonian. These are curves of constant torsion, with torsions $\tau=1$ and $2$ for the left and right respectively. They are comprised of helices, and are colored to demonstrate which part of the curve corresponds to which part of the pulse in the top part of the figure.}
	\label{fig:cc}
\end{figure}

Consider the case where $|E_1-E_2|\ll t_{\text{pulse}}^{-1}$, where $t_{\text{pulse}}$ is a time scale corresponding to the total length of a pulse. Then since $\kappa_3$ is proportional to $|E_1-E_2|$, it will be small, and thus the entire curve will lie close to a 3-dimensional subspace of the 6-dimensional space. Then using the 3D formalism inside this subspace will correct curves to leading order in $|E_1-E_2|t_{\text{pulse}}$. This happens because the 3D formalism is designed to correct against small $Z$ errors, so if $E_1$ is close to $E_2$, then $H_0$ can be treated as two copies of the same $2\times2$ Hamiltonian, and the $Z_1Z_2$ term can be included with the error in each case. This means that no real 2-qubit gates can be performed using this method, since these pulses correct against the $Z_1Z_2$ term used to create the interaction needed for 2-qubit gates in the first place. Thus, we will focus instead on the case where $E_1$ and $E_2$ differ significantly.

To demonstrate the utility of this formalism, we numerically derive a dynamically corrected gate that implements a $Z$ rotation on one qubit while canceling the effects of noise on both qubits. In the representation of the error as two 3D curves, this will require generating two curves with differing constant torsion, with the equal curvatures as functions of arclength. There has been much work deriving methods of generating closed curves of constant torsion \cite{WeinerAM1974,WeinerPAMS1977,CaliniPLA1999,BatesJG2013}; however, the analytic methods often used are not applicable here due to the additional constraints of generating two curves with equal curvatures. Thus, we approach the problem by numerically piecing together sets of helices to make two smooth closed curves, which corresponds to a series of square pulses canceling errors similar to \textsc{Supcode} \cite{WangNC2012}. Alternatively, in the 6-dimensional picture, higher dimensional generalizations of helices (curves with constant curvature coefficients) can be used. These will be five dimensional, since $\kappa_5$ is zero, due to $d\Omega/dt$ vanishing for a square pulse. Care must be taken for the step function transition between different values of $\Omega$, as $d\Omega/dt$ becomes a delta function at these points. At these transitions, $\Omega(t)$ can be treated as a steep constant slope over a short interval of time $\epsilon$, in the limit where $\epsilon\rightarrow 0$. In this case $\kappa_1$ through $\kappa_4$ will be finite, and thus will have no effect on the curve as $\epsilon\rightarrow 0$. $\kappa_5$ will have a delta function contribution, resulting in a rotation between $\hat{e}_5$ and $\hat{e}_6$ at the point along the curve corresponding to the transition. The exact angle of rotation $\phi$ corresponding to a step from $\Omega_1$ to $\Omega_2$ will be given by:
\begin{equation}
\phi=\int_{\Omega_1}^{\Omega_2}\frac{E_1E_2\sqrt{2(E_1^2+E_2^2)}}{\Omega^2(E_1^2+E_2^2)+2E_1^2E_2^2}\;d\Omega.
\end{equation}

\begin{figure}[!tb]
	\includegraphics[width=.6\columnwidth]{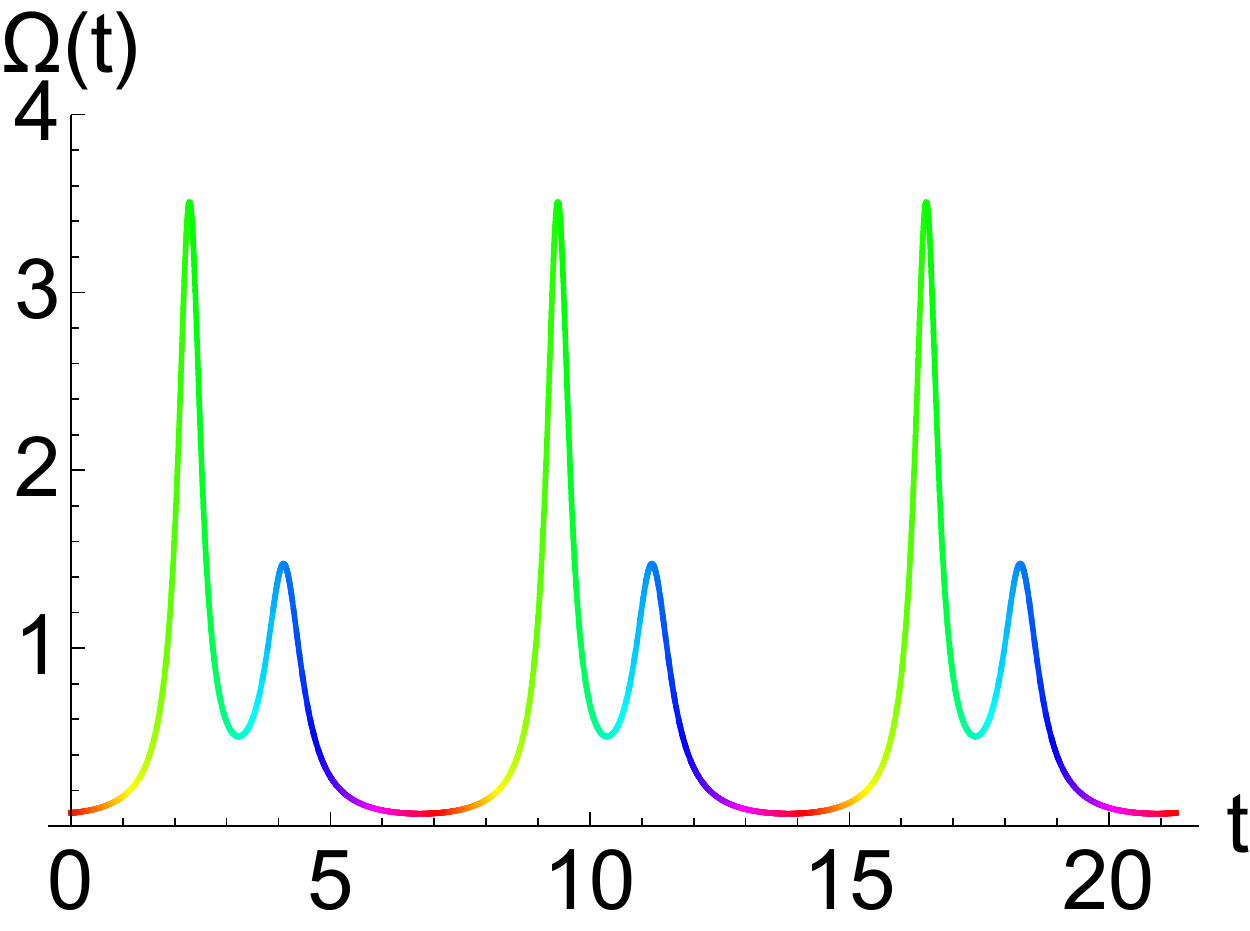}
	\includegraphics[width=.49\columnwidth]{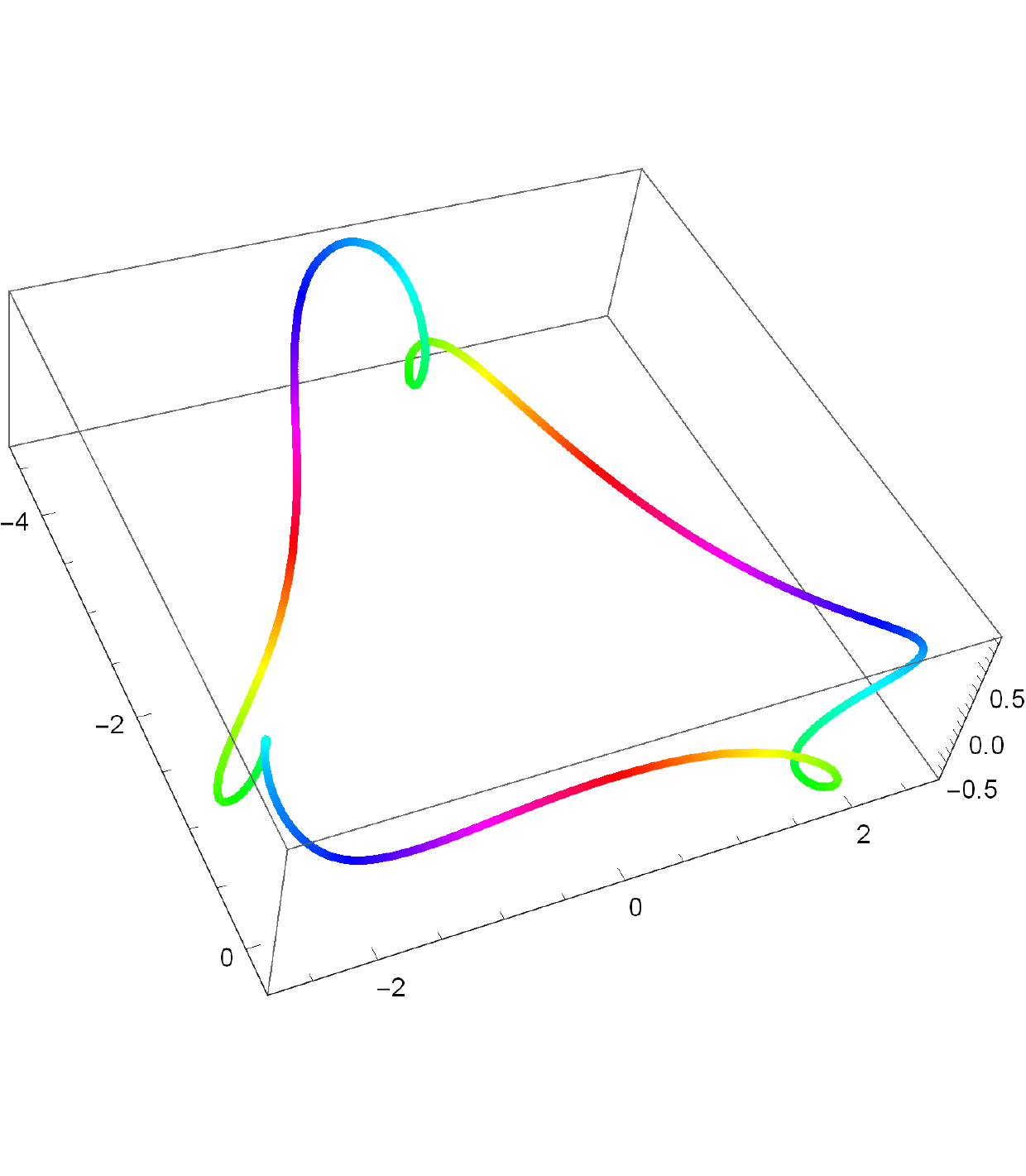}
	\includegraphics[width=.49\columnwidth]{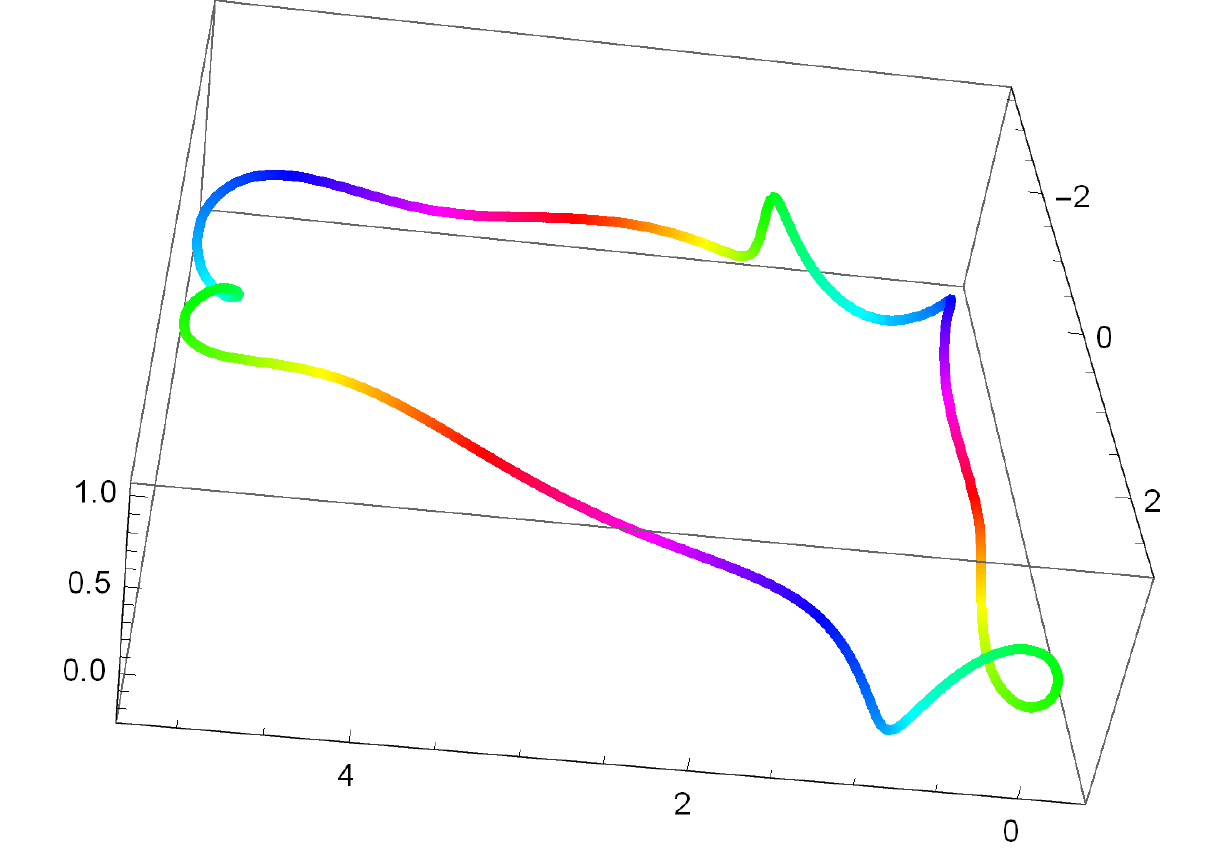}
	\caption{{\bf Top:} A smooth pulse implementing a $Z_1$ gate which dynamically cancels error to first order in $\delta H$ for the Hamiltonian given by eq. (\ref{eqn:h0}). This pulse is of the form given by eq. (\ref{eqn:smoothpulse}). {\bf Bottom:} Error curves for the two $2\times 2$ blocks of the Hamiltonian. These are curves of constant torsion, with torsions $\tau=1$ and $2$ for the left and right respectively. The curves are colored to illustrate which points correspond to which part of the pulse in the top part of the figure.}
	\label{fig:cs}
\end{figure}

\begin{figure}[!tb]
	\includegraphics[width=.49\columnwidth]{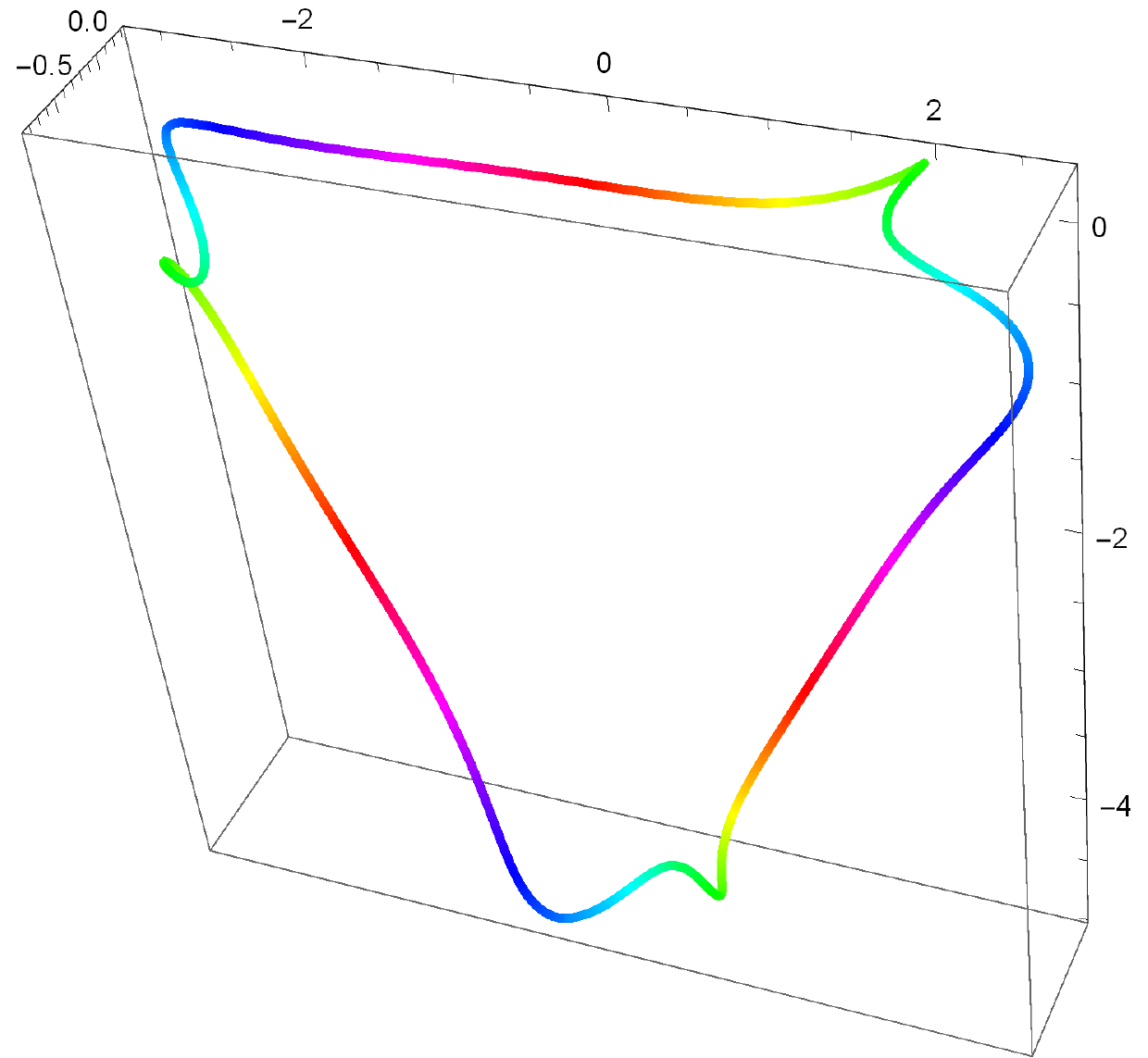}
	\includegraphics[width=.49\columnwidth]{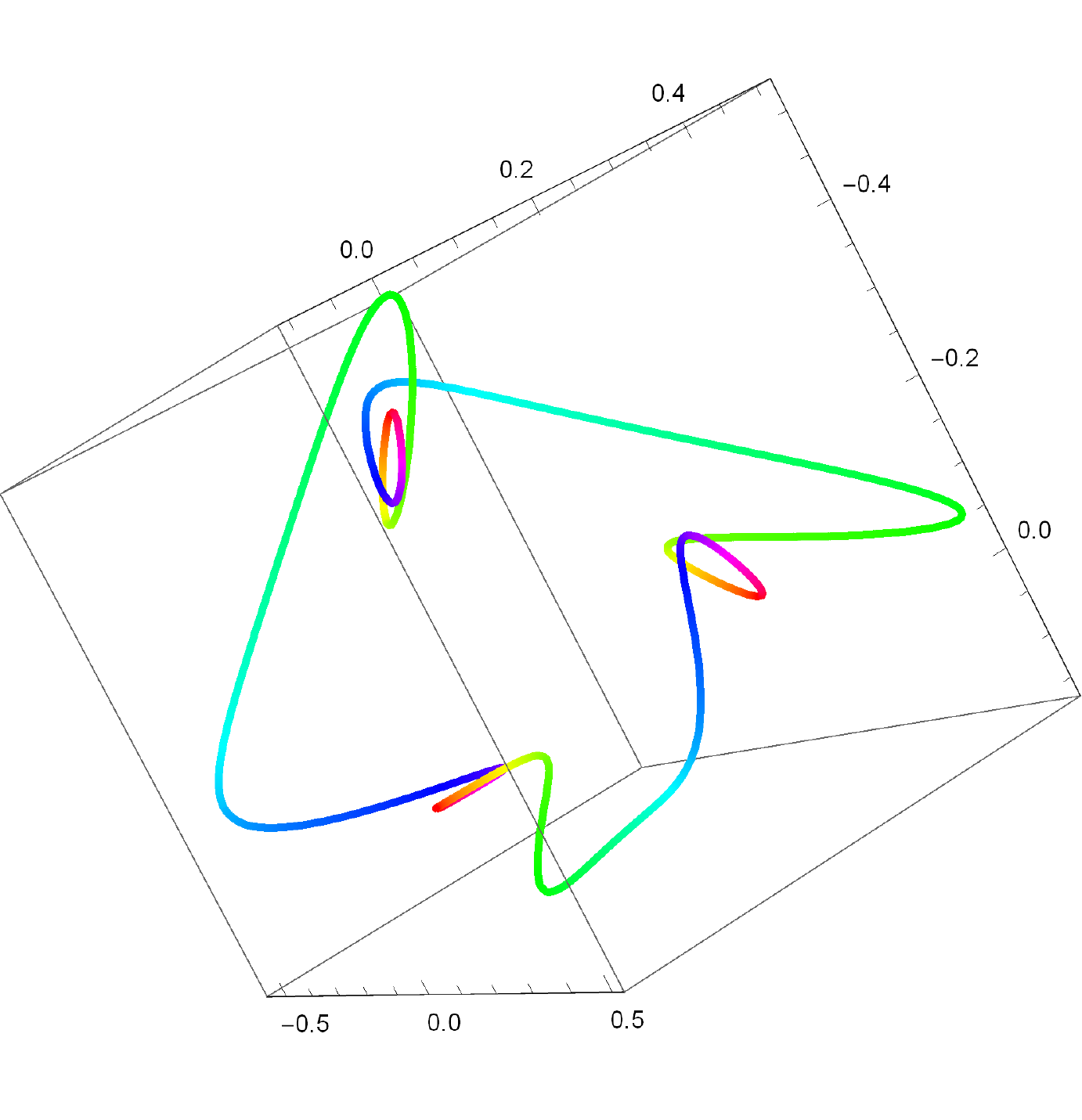}
	\caption{{\bf Left:} Projection of the 6D error curve for the pulse in Fig. \ref{fig:cs} onto the $\vec{e}_1(0)$, $\vec{e}_3(0)$, $\vec{e}_4(0)$ subspace. The two largest dimensions correspond to $\vec{e}_1(0)$ and $\vec{e}_4(0)$. {\bf Right:} Projection of the 6D curve onto the $\vec{e}_2(0)$, $\vec{e}_5(0)$, $\vec{e}_6(0)$ subspace. Note that the length of the curve in these dimensions is much smaller than in dimensions 1 and 4 on the left, and is comparable to dimension 3 on the left.}
	\label{fig:cs6}
\end{figure}

We choose to consider curves which are $n$-fold rotationally symmetric, as these curves are easier to visualize and require fewer parameters to represent. To this end, we consider one curve segment where the Frenet-Serret frame at one endpoint is equal to the Frenet-Serret frame at the other endpoint after having undergone a relative $2k\pi/n$ rotation, for any integer $k$ coprime to $n$. Any curve segment such as this will produce a closed curve when repeated $n$ times, provided that the displacement vector between endpoints lies within the plane of rotation. For our numerics, we consider 3-fold symmetric curves, which amounts to choosing a periodic pulse with period equal to one third of the total pulse length. Then we numerically adjust the parameters corresponding to the legnths and curvatures of the helices until the conditions on the displacement vector and the Frenet-Serret frames at the endpoints are met. Fig. \ref{fig:cc} shows an example of a square pulse derived in this way, with $E_2=2E_1$.

In experiments, square pulses cannot be exactly performed, since pulse generators have bandwidth and pulse rise-time limitations. However, we can numerically search for a smooth pulse similar in shape to the square pulse already obtained. We do this by choosing a pulse shape of the following form:
\begin{equation}
\Omega(t)=c_0+\frac{c_1}{1+a_1^2\sin^2(\pi t/t_p+\phi_1)}+\frac{c_2}{1+a_2^2\sin^2(\pi t/t_p+\phi_2)}.
\label{eqn:smoothpulse}
\end{equation}
The form is meant to approximate a Lorentzian pulse, except that it is periodic with period $t_p$. The curves can then be numerically generated, and the parameters in Eq. (\ref{eqn:smoothpulse}) adjusted until closed curves are obtained. Because we use more parameters than the dimension of the space, solutions which produce closed curves are not necessarily unique. Uniqueness is not essential here since all we need is a solution providing dynamical decoupling, and the fact that there may be other solutions is not a problem. If finding solutions inside a desired parameter regime is difficult, a third Lorentzian can be added to the pulse to allow for more parameters. Adding more parameters should simplify the process of finding a possible solution as long as we do not insist on unique solutions; however the shape of the resulting pulse will be more complicated as more parameters are used. Using this method, we find the pulse and curves shown in Fig. \ref{fig:cs}. In Fig. \ref{fig:cs6}, we plot the error curve in the 6-dimensional representation of the same pulse by projecting it into two 3D subspaces. We see that it retains its 3-fold rotational symmetry. In two of these dimensions, $\vec{e}_1(0)$ and $\vec{e}_4(0)$, the curve covers much more distance than in the other four.

\begin{figure}[!htb]
	\includegraphics[width=\columnwidth]{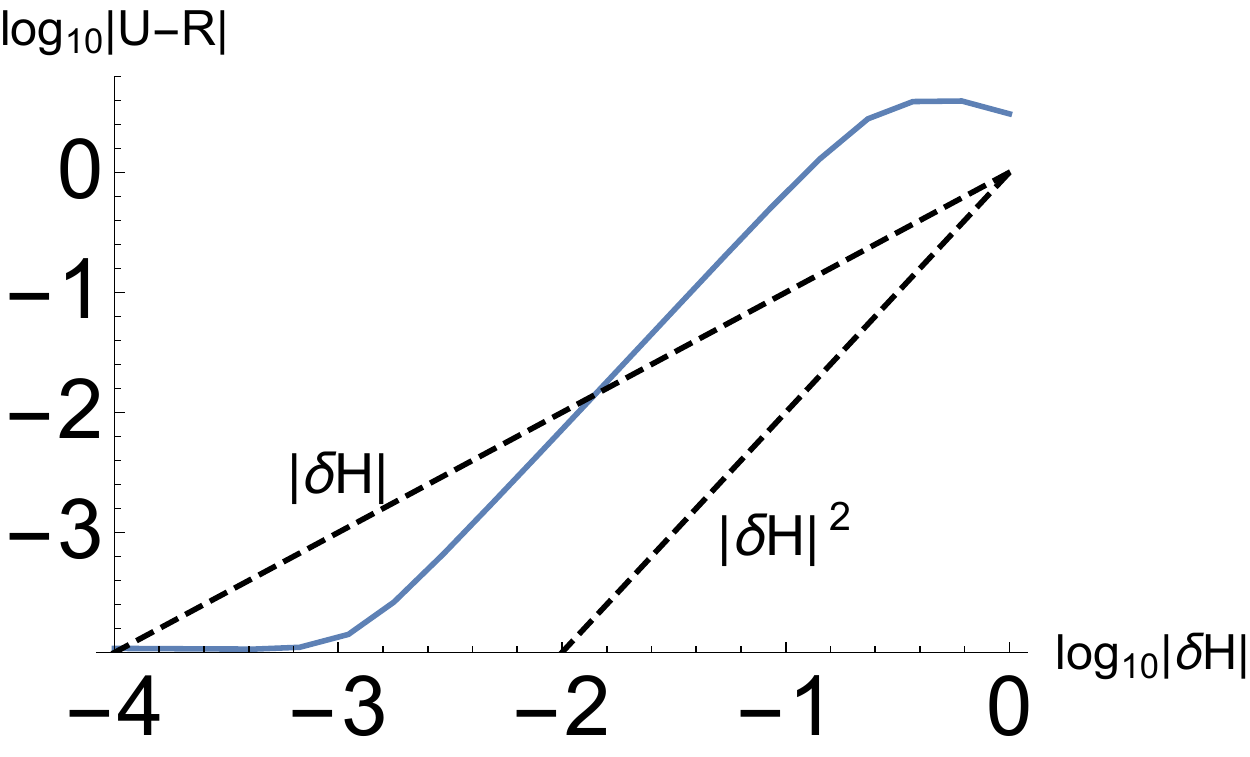}
	\caption{Infidelity versus noise strength of the pulse given by Fig. \ref{fig:cs} (solid blue curve). The dashed lines show linear and quadratic scaling with noise strength. It is clear that gate infidelity is consistent with quadratic scaling, indicating that first-order noise errors have been canceled.}
	\label{fig:infplot}
\end{figure}

By looking at each $2\times 2$ block in the block-diagonal matrix individually, we can use the same method as with the single-qubit case to determine the gates performed. For symmetric pulses like the ones we generated, this will result in $\vec{e}_1(t_{\text{pulse}})=\vec{e}_1(0)$, and similarly for $\vec{e}_2$ and $\vec{e}_3$, where these vectors belong to the 3-dimensional spaces corresponding to each block. Thus, these symmetric pulses can only perform identity operations in each of the 2$\times$2 blocks. These identity operations can have a different relative sign, resulting in a $Z_1$ gate, as is the case with the pulse shown in Fig. \ref{fig:cs}. Removing the symmetry requirement will generate other gates. In order to demonstrate that these pulses do properly correct against error, we numerically calculate the infidelity of the gate using the full noisy Hamiltonian for noise strengths across several orders of magnitude. Here, we consider a single quasistatic $Z_2$ noise source, and vary its strength, plotting the resulting infidelity versus the noise strength in Fig. \ref{fig:infplot}. We define infidelity as $|U-R|$, where $U$ is the noisy pulse, and $R$ is the ideal pulse (in this case $Z_1$). We see that the infidelity scales as the square of the noise strength, indicating that the gate cancels error to first order in $|\delta H|$. Thus, our proposed geometric scheme does correct for noise in the Ising 2-qubit gate operations.

\section{Conclusion}

We have shown how to generalize the geometric formalism for producing dynamically corrected single-qubit gates to multiqubit systems. Like the single-qubit case, the cumulative first-order error can be represented as a curve in Euclidean space. Distance along the curve corresponds to the elapsed time from the beginning of the pulse, and the strength of the driving fields can be related to the curvature coefficients at each point on the curve. Critically, using these curvature coefficients circumvents the need to evaluate the time-ordered exponential of the Hamiltonian, which in general cannot be done except in very special cases. We presented equations which show how to calculate these curvature coefficients in terms of the time derivatives and commutators of the Hamiltonian $H_0$ and noise source $\delta H$. We used this formalism to numerically derive dynamically corrected gates for a two-qubit Hamiltonian that is relevant for Ising-coupled qubits, as relevant for superconducting transmons and singlet-triplet spin qubits, and demonstrated that the pulses cancel first-order error by computing the infidelity of the gate as a function of the noise strength.

While this formalism provides a good starting point for deriving dynamically corrected gates for multiqubit systems, there are still several challenges. In particular, it is usually the case that only a few terms in the Hamiltonian can be controlled dynamically. In this case, we need to restrict to the subset of curves that produce Hamiltonians of the desired form. This issue arises when the number of control fields is less than the number of generalized curvatures, in which case we need to find curves for which some of the generalized curvatures remain constant. Note that this is an issue even in the single-qubit case if only one control field is present. In this circumstance, one needs to restrict to closed curves of constant torsion. How to do this in general for multiple curvatures is an important open question for future work.

\acknowledgments

We thank Fei Zhuang for useful discussions. E.B. acknowledges support from the Office of Naval Research (Grant No. N00014-17-1-2971). This work is supported by the Laboratory for Physical Sciences.

\end{document}